\documentclass{revtex4}
\usepackage{amssymb}
\usepackage{amsfonts}
\usepackage{amsmath}

\input{tcilatex}

\begin{document}

\title{On the nonlinear dynamics of a position-dependent mass-driven
Duffing-type oscillator: Lagrange and Newton equations' equivalence}
\author{Omar Mustafa}
\email{omar.mustafa@emu.edu.tr}
\affiliation{Department of Physics, Eastern Mediterranean University, G. Magusa, north
Cyprus, Mersin 10 - Turkey,\\
Tel.: +90 392 6301378; fax: +90 3692 365 1604.}

\begin{abstract}
\textbf{Abstract:} Using a generalized coordinate along with a proper
invertible coordinate transformation, we show that the Euler-Lagrange
equation used by Bagchi et al. \cite{16} is in clear violation of the
Hamilton's principle. We also show that Newton's equation of motion they
have used is not in a form that satisfies the dynamics of position-dependent
mass (PDM) settings.. The equivalence between Euler-Lagrange's and Newton's
equations is now proved and documented through the proper invertible
coordinate transformation and the introduction of a new PDM byproducted
reaction-type force. The total mechanical energy for the PDM is shown to be
conservative (i.e., $dE/dt=0$, unlike Bagchi et al.'s \cite{16} observation).


\end{abstract}

\maketitle

The growing interests and/or research developments in the position-dependent
mass (PDM) quantum mechanical systems described (mainly) by the von Roos
Hamiltonian \cite{1} (see also the sample of references \cite%
{2,3,4,5,6,7,8,9,10,11,12} and related references cited therein) have
inspired the relatively recent\ and rapid research attention in the PDM for
classical mechanical systems (cf, e.g., \cite{12,13,14,16,16,17}). In fact,
the interest in the PDM classical particles dates back to 1974 through the
Mathews and Lakshmanan \cite{18} study of the equation of motion%
\begin{equation}
\left( 1+\xi x^{2}\right) \ddot{x}-\xi x\,\dot{x}^{2}+\omega _{\circ }^{2}x=0%
\text{ };\;\xi \in 
\mathbb{R}
,
\end{equation}%
where an overhead dote indicates a time derivative. This equation
corresponds to a nonlinear oscillator, exhibiting amplitude-dependent simple
harmonic oscillations \cite{19}, with the Lagrangian%
\begin{equation}
L=\frac{1}{2}\left( \frac{\dot{x}^{2}-\omega _{\circ }^{2}x^{2}}{1+\xi x^{2}}%
\right)
\end{equation}%
and the canonical linear momentum%
\begin{equation}
p=m\left( x\right) \,\dot{x}\,\text{; \ }m\left( x\right) =\frac{1}{1+\xi
x^{2}}.
\end{equation}

In their attempt to reproduce the Lagrange equation in (1), Bagchi et al. 
\cite{16} have proposed that in the absence of any external force term the
Newton's equation of motion with PDM gets modified to%
\begin{equation}
m\left( x\right) \ddot{x}+m^{^{\prime }}\left( x\right) \dot{x}^{2}=0,
\end{equation}%
where the prime indicates spatial derivative. Which when compared with (1)
would suggest a PDM function in the form of%
\begin{equation}
m\left( x\right) =\frac{1}{\sqrt{1+\xi x^{2}}},
\end{equation}%
on ignoring, of course, the presence of the harmonic term $\omega _{\circ
}^{2}x$ to effect such a comparison. They have also recollected Cruz et al's
work \cite{14} and used%
\begin{equation}
\frac{d}{dt}\left( \frac{\partial L}{\partial \dot{x}}\right) -\frac{%
\partial L}{\partial x}=\tilde{R}\,;\text{ }\tilde{R}=-\frac{1}{2}%
m^{^{\prime }}\left( x\right) \dot{x}^{2},
\end{equation}%
where $\tilde{R}$ is a reaction thrust (as named by Cruz et al. \cite{14}).
This would make their Lagrange's and Newton's equations of motion equivalent
and consistent. However, this is an improper approach for the PDM-settings
in the classical mechanical framework.

A priory, the Lagrange's equation,\ used by Cruz et al \cite{14}, in (6) is
a clear violation of the textbook Hamilton's least action principle, one of
the most profound results in physics. Yet, one would notice that the
Newton's equation in (4) is based on the conservation of the linear
momentum, $dp/dt=\dot{p}=0$, which is an improper approach for PDM settings.
Mazharimousavi and Mustafa \cite{17} have very recently shown that the
quasi-linear momentum (as maned therein)%
\begin{equation}
\Pi \left( x,\dot{x}\right) =\sqrt{m\left( x\right) }\dot{x}
\end{equation}%
is the conserved quantity (i.e., $\Pi \left( x,\dot{x}\right) =\Pi
_{0}\left( x_{0},\dot{x}_{0}\right) $ and $\dot{\Pi}\left( x,\dot{x}\right)
=0$ , where $x_{0}$ and $\dot{x}_{0}$ are the initial position and initial
velocity of the PDM, respectively) and not the linear momentum (i.e., $%
p\left( x,\dot{x}\right) \neq p_{0}\left( x_{0},\dot{x}_{0}\right) $, and $%
\dot{p}\left( x,\dot{x}\right) \neq 0$). In this communication, we fix this
issue through the following arguments.

Let us consider a classical particle with \emph{"unit mass"} \ in the
generalized coordinate $q$ moving with velocity $\dot{q}$ in a force-free
field, $V\left( q\right) =0$. Then the corresponding Lagrangian $L\left( q,%
\dot{q}\right) $ is given by 
\begin{equation}
L\equiv L\left( q,\dot{q}\right) =\frac{1}{2}\dot{q}^{2},
\end{equation}%
and satisfies the Euler-Lagrange's equation%
\begin{equation}
\frac{d}{dt}\left( \frac{\partial L}{\partial \dot{q}}\right) -\frac{%
\partial L}{\partial q}=0.
\end{equation}%
Under such assumptions, Eq.(9) would that $\ddot{q}=0$. Which is in exact
accord with Newton's law of motion for a free classical particle moving in
the generalized coordinate $q$ with a conserved generalized momentum%
\begin{equation*}
p_{q}=\frac{dq}{dt}=0\Longrightarrow \dot{q}=\dot{q}_{0},
\end{equation*}%
where $\dot{q}_{0}$ is the generalized initial momentum. Now, let the
coordinate transformation%
\begin{equation}
q\equiv q\left( x\right) =\int^{x}\sqrt{f\left( u\right) }du\Longrightarrow
q^{\prime }\left( x\right) =\sqrt{f\left( x\right) }\Longrightarrow \dot{q}%
\left( x\right) =\dot{x}\sqrt{f\left( x\right) }
\end{equation}%
represent a mapping from the coordinate $q$ onto the coordinate $x$. As long
as our choice of the coordinates is invertible (i.e., $det\left( \partial
x_{i}/\partial q_{i}\right) \neq 0$, in general, which is the case under
consideration here) then one can easily show that the Euler-Lagrange
equation in (9) reads%
\begin{equation}
\left[ \frac{d}{dt}\left( \frac{\partial L}{\partial \dot{x}}\right) -\frac{%
\partial L}{\partial x}\right] \left( \frac{\partial x}{\partial q}\right)
=0\Longrightarrow \frac{d}{dt}\left( \frac{\partial L}{\partial \dot{x}}%
\right) -\frac{\partial L}{\partial x}=0.
\end{equation}%
This in fact follows immediately from the Hamilton's least action principle.
The Euler-Lagrange equations are known to be coordinate invariant for they
assume the same form in all coordinate systems, provided that the
coordinates invertibility is secured. As such, the Lagrangian $L\left( q,%
\dot{q}\right) $ in (8) would transform into%
\begin{equation}
L\left( x,\dot{x}\right) =\frac{1}{2}f\left( x\right) \dot{x}^{2}
\end{equation}%
with the corresponding Euler-Lagrange equation%
\begin{equation}
f\left( x\right) \ddot{x}+\frac{1}{2}f^{^{\prime }}\left( x\right) \dot{x}%
^{2}=0.
\end{equation}

One may now go backwards and start with the Langrangian $L\left( x,\dot{x}%
\right) =\frac{1}{2}m\left( x\right) \dot{x}^{2}$ of a PDM particle $m\left(
x\right) =$ $f\left( x\right) $ moving in a force-free field, $V\left(
x\right) =0$, and use (11) to obtain%
\begin{equation}
m\left( x\right) \ddot{x}=-\frac{1}{2}m^{^{\prime }}\left( x\right) \dot{x}%
^{2}\Longrightarrow \sqrt{m\left( x\right) }\dot{x}=\sqrt{m\left(
x_{0}\right) }\dot{x}_{0}\Longrightarrow \Pi \left( x,\dot{x}\right) =\Pi
_{0}\left( x_{0},\dot{x}_{0}\right) .
\end{equation}%
This result would not only support our thesis in \cite{17} on the
conservation of the quasi-linear momentum (i.e., $\dot{\Pi}\left( x,\dot{x}%
\right) =0$) but also suggests the amendment that has to made for Newton's
equation of motion (4) for PDM used by Bagchi et al. \cite{16} and by Cruz
et al. \cite{14} in the absence of any external force term. In what follows
we shall see that the "any external force" terminology holds true only for
constant mass settings.

In this regard, we may express the linear momentum in terms of the
quasi-linear momentum%
\begin{equation}
p=m\left( x\right) \dot{x}=\sqrt{m\left( x\right) }\Pi \left( x,\dot{x}%
\right) ,
\end{equation}%
and cast Newton's equation of motion as%
\begin{equation}
F_{ext}\left( x,\dot{x}\right) =F_{ext}\left( x\right) +R_{PDM}\left( x,\dot{%
x}\right) =\frac{dp}{dt}\text{ ,}
\end{equation}%
where $F_{ext}\left( x\right) =-\partial V\left( x\right) /\partial x=0$ is
the set of potential energy driven external forces, and $R_{PDM}\left( x,%
\dot{x}\right) $ represents any feasible PDM-byproducted reaction-type force
(of course, if it exists at all). To find out such a PDM-byproducted
reaction force, $R_{PDM}\left( x,\dot{x}\right) $, one would naturally find%
\begin{equation}
\frac{dp}{dt}=\frac{d}{dt}\left[ m\left( x\right) \dot{x}\right] =m\left(
x\right) \ddot{x}+m^{^{\prime }}\left( x\right) \dot{x}^{2},
\end{equation}%
and%
\begin{equation}
R_{PDM}\left( x,\dot{x}\right) =\frac{dp}{dt}=\frac{d}{dt}\left[ \sqrt{%
m\left( x\right) }\Pi \left( x,\dot{x}\right) \right] =\frac{1}{2}\frac{%
m^{^{\prime }}\left( x\right) }{\sqrt{m\left( x\right) }}\dot{x}\Pi \left( x,%
\dot{x}\right) =\frac{1}{2}m^{^{\prime }}\left( x\right) \dot{x}^{2}.
\end{equation}%
When (17) and (18) are substituted in the Newton's equation of motion in
(16), along with $F_{ext}\left( x\right) =0$, one obtains%
\begin{equation}
\frac{1}{2}m^{^{\prime }}\left( x\right) \dot{x}^{2}=m\left( x\right) \ddot{x%
}+m^{^{\prime }}\left( x\right) \dot{x}^{2}\Longrightarrow m\left( x\right) 
\ddot{x}+\frac{1}{2}m^{^{\prime }}\left( x\right) \dot{x}^{2}=0
\end{equation}%
which is in exact accord with the Euler-Lagrange result in (14). Obviously,
the equivalence between the Euler-Lagrange's and Newton's equations is now
documented through the above "good" invertible coordinate transformation
(i.e., $\partial x/\partial q\neq 0\neq \partial q/\partial x$) and the
introduction of the new PDM-byproducted reaction-type force $R_{PDM}\left( x,%
\dot{x}\right) $ into Newton$^{\prime }$s law of motion. Hereby, we may
safely conclude that the PDM setting is nothings but a manifestation of some
"good" invertible coordinate transformation that leaves the corresponding
Euler-Lagrange equation invariant.

Under such textbook documentation settings, one would write the PDM
Lagrangian as%
\begin{equation}
L=T-V=\frac{1}{2}m\left( x\right) \dot{x}^{2}-V\left( x\right) \,,\text{ }%
F_{ext}\left( x\right) =-\frac{\partial V\left( x\right) }{\partial x}\neq 0,
\end{equation}%
with the corresponding Euler-Lagrange equation%
\begin{equation}
m\left( x\right) \ddot{x}+\frac{1}{2}m^{^{\prime }}\left( x\right) \dot{x}%
^{2}+\frac{\partial V\left( x\right) }{\partial x}=0.
\end{equation}%
The Mathews and Lakshmanan \cite{18} equation of motion in (1) is obtained
if a PDM particle with $m\left( x\right) =\left( 1+\xi x^{2}\right) ^{-1}$
is subjected to move in an oscillator potential $V\left( x\right) =\frac{1}{2%
}m\left( x\right) \omega _{\circ }^{2}x^{2}$ or a hypothetical potential $%
V\left( x\right) =-\frac{1}{2}m\left( x\right) \omega _{\circ }^{2}/\xi $ 
\cite{20} (if such potential exists at all). Both potentials yield%
\begin{equation}
m\left( x\right) \ddot{x}-m\left( x\right) ^{2}\xi x\dot{x}^{2}+m\left(
x\right) ^{2}\omega _{\circ }^{2}x=0.
\end{equation}%
Therefore, in the presence of an external periodic force with additional
damping term and a quartic potential, the result in (22) would immediately
suggest the following amendment to Eq. (11) of Bagchi et al.'s \cite{16} to
read%
\begin{equation}
m\left( x\right) \ddot{x}-m\left( x\right) ^{2}\xi x\dot{x}^{2}+m\left(
x\right) ^{2}\omega _{\circ }^{2}x+\lambda x^{3}+\alpha \dot{x}=f\cos \omega
t.
\end{equation}%
Which for a constant unit mass (i.e., $\xi =0$) recovers a forced, damped
Duffing oscillator. Yet, with $\dot{x}=y$ Eq.(12) of Bagchi et al. \cite{16}
should (taking $\dot{z}=\omega $) read%
\begin{equation}
\dot{y}=\frac{\xi x\dot{x}^{2}-\omega _{\circ }^{2}x}{1+\xi x^{2}}+\left(
1+\xi x^{2}\right) \left[ f\cos z-\lambda x^{3}-\alpha y\right] .
\end{equation}

Moreover, one can show that the time rate of change of the total mechanical
energy $E=T+V$ is given by%
\begin{equation}
\frac{dE}{dt}=\left[ m\left( x\right) \ddot{x}+\frac{1}{2}m^{^{\prime
}}\left( x\right) \dot{x}^{2}+\frac{\partial V\left( x\right) }{\partial x}%
\right] \dot{x}\Longrightarrow \frac{dE}{dt}=0,
\end{equation}%
indicating that the total mechanical energy is conserved (unlike equation
(10) of Bagchi et al. \cite{16}) and is the constant of motion. Therefore,
the PDM-buproducted reaction force $R_{PDM}\left( x,\dot{x}\right) $ turns
out to be a conservative force with explicit dependence on both position and
velocity. Should we assume that $m\left( x\right) $ and $m^{^{\prime
}}\left( x\right) $ are both positive valued functions, then $R_{PDM}\left(
x,\dot{x}\right) $ would act in the direction of the time rate of change of
the linear momentum (i.e., $R_{PDM}\left( x,\dot{x}\right)
=dp/dt=m^{^{\prime }}\left( x\right) \dot{x}^{2}/2$). The consequences of
the very existence of such a force are discussed by Mazharimousavi and
Mustafa \cite{17} through the use of the Euler-Langrange's equation (19) and
the related examples reported therein.

For the sake of fairness and/or completeness, if one assumes that a PDM
particle $m\left( x\right) =1/\sqrt{1+\xi x^{2}}$ is moving under the
influence of the set of mass-independent forces (as apparently suggested by
Bagchi et al. \cite{16}), only then Eq.(12) of Bagchi et al. would need a
multiplicity of order 1/2 in the first term of $\dot{y}$ (as a consequence
of our proposed amendment of Newton's low in (19)). This should not affect
their numerical study of the propagation dynamics and the sensitivity of the
PDM-index $\xi $ to enhance phase transition from a limit cycle mode to a
chaotic regime and the initiation of the complicated nature of bifurcation,
etc.


\begin{thebibliography}{99}
\bibitem{1} O. Von Roos, Phys. Rev. \textbf{B 27} (1983) 7547.

\bibitem{2} A. de Souza Dutra, C A S Almeida, Phys Lett. \textbf{A 275}
(2000) 25.

\bibitem{3} S. H. Mazharimousavi, J Phys \textbf{A}: Math. Theor.\textbf{41 (%
}2008\textbf{) }244016.

\bibitem{4} O. Mustafa, S. H. Mazharimousavi, Phys. Lett. \textbf{A 358}
(2006) 259.

\bibitem{5} A. D. Alhaidari, Phys. Rev. \textbf{A 66} (2002) 042116.

\bibitem{6} B. Bagchi, A. Banerjee, C. Quesne, V. M. Tkachuk, J. \ Phys. 
\textbf{A}: Math. Gen. \textbf{38} (2005) 2929.

\bibitem{7} O. Mustafa, S. H. Mazharimousavi, Phys. Lett. \textbf{A 357}
(2006) 295.

\bibitem{8} O. Mustafa, J Phys \textbf{A}: Math. Theor. \textbf{44 (}2011%
\textbf{) }355303.

\bibitem{9} B. Bagchi, P. Gorain, C. Quesne and R. Roychoudhury, Mod. Phys.
Lett. \textbf{A 19} (2004) 2765.

\bibitem{10} O. Mustafa, S. H. Mazharimousavi, Int. J. Theor. Phys \ \textbf{%
46} (2007) 1786.

\bibitem{11} S. Cruz y Cruz, O Rosas-Ortiz, J Phys \textbf{A}: Math. Theor. 
\textbf{42} (2009) 185205.

\bibitem{12} S. Cruz y Cruz, J. Negro and L.M. Nieto, Phys. Lett. \textbf{A} 
\textbf{369,} (2007) 400.

\bibitem{13} S. Cruz y Cruz, J. Negro and L.M. Nieto, Journal of Physics:
Conference Series \textbf{128, }(2008)\textbf{\ }012053.

\bibitem{14} S. Cruz y Cruz and O. R. Ortiz, SIGMA \textbf{9, }(2013)\textbf{%
\ }004.

\bibitem{15} S. Ghosh and S. K. Modak, Phys. Lett. A \textbf{373,} (2009)
1212.

\bibitem{16} B. Bagchi, S. Das, S. Ghosh and S. Poria, J. Phys. \textbf{A}:
Math. Theor. \textbf{46,} (2013) 032001.

\bibitem{17} S. H. Mazharimousevi, O. Mustafa, arXiv:1208.1095, to appear in
Physica Scripta.

\bibitem{18} P. M. Mathews, M. Lakshmanan, Quart. Appl. Math. \textbf{32,
(1974) }215.

\bibitem{19} A. Venkatesan, M. Lakshmanan, Phys. Rev. \textbf{E} \textbf{55}
(1997) 5134.

\bibitem{20} Z. E. Musielak, J. Phys. \textbf{A}: Math. Theor. \textbf{41,}
(2008) 055205.
\end{thebibliography}
\end{document}